\documentstyle[aps,prl,epsfig]{revtex}
\include{confidential}

\begin{document}

\title{Nuclear Effects on $R = \sigma_L / \sigma_T$ in
                Deep-Inelastic Scattering \\
      }

\vspace{5 mm}

\author{
\centerline {\it The HERMES Collaboration}\medskip \\
K.~Ackerstaff$^{5}$, 
A.~Airapetian$^{33}$, 
N.~Akopov$^{33}$,
I.~Akushevich$^{6}$,
M.~Amarian$^{25,28,33}$, 
E.C.~Aschenauer$^{6,13,25}$, 
H.~Avakian$^{10}$, 
R.~Avakian$^{33}$, 
A.~Avetissian$^{33}$, 
B.~Bains$^{15}$
% ...
}

\address{
% ...
$^5$DESY, Deutsches Elektronen Synchrotron, 22603 Hamburg, Germany\\
$^6$DESY Zeuthen, 15738 Zeuthen, Germany\\
% ...
$^{10}$Istituto Nazionale di Fisica Nucleare, Laboratori Nazionali di Frascati, 00044 Frascati, Italy\\
% ...
$^{13}$Department of Subatomic and Radiation Physics, University of Gent, 9000 Gent, Belgium\\
% ...
$^{15}$Department of Physics, University of Illinois, Urbana, Illinois 61801, USA\\
% ...
$^{25}$Nationaal Instituut voor Kernfysica en Hoge-Energiefysica (NIKHEF), 1009 DB Amsterdam, The Netherlands\\
% ...
$^{28}$Istituto Nazionale di Fisica Nucleare, Sezione Sanit\`a and Physics Laboratory, Istituto Superiore di Sanit\`a, 00161 Roma, Italy\\
% ...
$^{33}$Yerevan Physics Institute, 375036, Yerevan, Armenia
}
%%%%%%%%%%%%%%%%%%%%%%%%%%%%%%%%%%%%%%%%%%%%%%%%%%%%%%%%%%%%%%%%%%%%%%%%%%%%%%%

\date{\today}
\maketitle

\begin{abstract}
Cross section ratios for deep-inelastic scattering from 
$^{14}$N and $^3$He with respect to $^2$H have been measured
by the HERMES experiment at DESY using a 27.5 GeV positron beam.
The data cover a range in the Bjorken scaling
variable $x$ between 0.013 and 0.65, while the negative squared 
four-momentum transfer $Q^2$ varies from 0.5 to 15 GeV$^2$.
The data are compared to measurements performed by NMC, E665, and SLAC
on $^4$He and $^{12}$C, and are found to be different for 
$x <$ 0.06 and $Q^2 <$ 1.5 GeV$^2$. 
The observed difference is attributed to 
an $A$-dependence of the ratio $R = \sigma_L / \sigma_T$
of longitudinal to transverse 
deep-inelastic scattering cross sections
at low $x$ and low $Q^2$. 
\end{abstract}

\medskip

\vspace{1cm}

\centerline{PACS numbers: 13.60.Hb, 13.60.-r, 24.85.+p, 12.38.-t}

\twocolumn
 
The energy scales relevant to deep-inelastic lepton nucleon scattering 
(multi-GeV) greatly differ from
those relevant to the atomic nucleus (multi-MeV). Hence, it came as
a surprise that the structure function $F_2^N(x)$, which in the
Quark-Parton Model
represents the quark momentum distribution inside the nucleon,
was found to depend on the mass $A$ of the atomic nucleus \cite{emc-effect}. 
This phenomenon is known as the {\it EMC effect} at large values of the
Bjorken scaling variable $x$, i.e. $x > 0.1$, and as {\it shadowing} 
at lower values of $x$ \cite{Arne94}.

With $F_2(x)$ found to be $A$-dependent, it is relevant to investigate
whether this dependence is the same for its longitudinal 
and transverse components, $F_L(x)$ and $F_1(x)$.
The latter two structure functions are related to $F_2(x)$ via
$F_L(x) = (1 + Q^2/\nu^2) F_2(x) - 2 x F_1(x)$
with $Q^2$ the negative of the four-momentum transfer squared $q^2$, 
$\nu$ the energy transfer, $x = Q^2 / 2 M \nu$ and $M$ the nucleon
mass. A possible difference of the $A$-dependence of 
$F_L(x)$ and $F_1(x)$ can be investigated by measuring
the ratio of longitudinal to transverse deep-inelastic
scattering (DIS) cross sections 
$R = \sigma_L / \sigma_T = F_L(x) / 2 x F_1(x)$
for various nuclear targets. 

Theoretically, a possible $A$-dependence of $R$ has been suggested
by several authors. In ref. \cite{Mila94} the  Fermi motion of
the nucleons is seen to enhance higher-twist effects, which will lead to
an enhancement of $F_L(x)$ at low values of $x$ and $Q^2$. It has
also been argued \cite{Gou96} that an increase of the nuclear gluon
distribution may
lead to an enhancement of $R$. On the other hand, in ref. \cite{Niko91}
it is suggested that nuclear shadowing might be 
different for the longitudinal and transverse DIS cross sections.
The predicted size and ($x$,$Q^2$)-dependence of these effects are all
different. However, no experimental evidence for
an $A$-dependence of $R$ has been found to date 
\cite{Amau92,Arne96,Dasu94,Tao96}.

\begin{figure} [t]
\begin{center}
\includegraphics[width=0.47\textwidth]{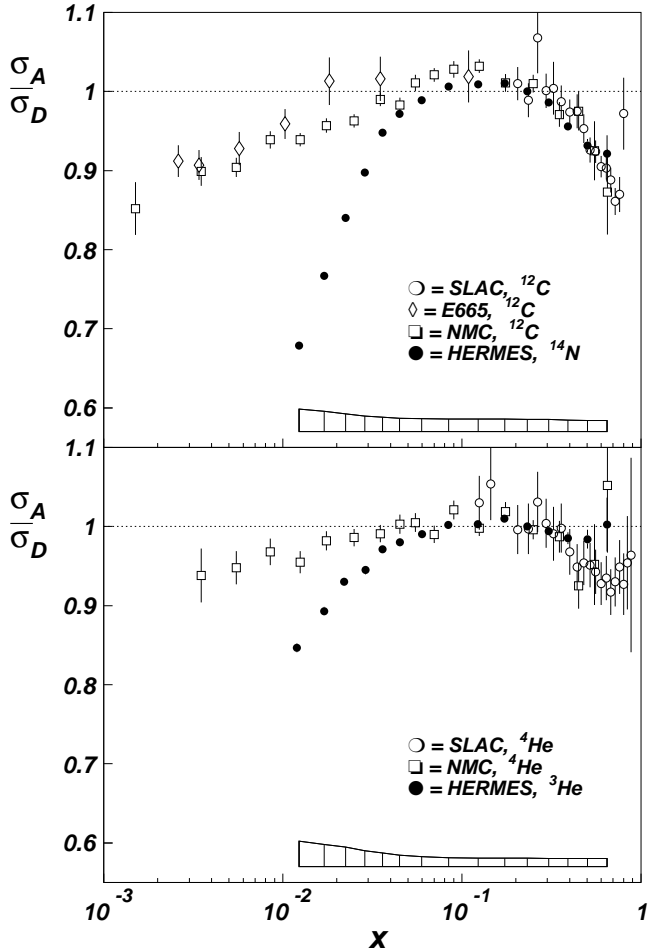}
\epsfxsize=\hsize
\end{center}
\caption{
Ratio of cross sections of
inclusive deep-inelastic lepton scattering from nucleus
$A$ and $D$ versus $x$.
The error bars of the HERMES measurement represent the statistical
uncertainties, the systematic uncertainty of the HERMES data is given by the
error band.
The error bars of the NMC, E665, and SLAC data are given by
the quadratic sum of the
statistical and systematic uncertainties.
\label{fig:14n-x} }
\end{figure}

In this Letter we present data from the HERMES experiment
on the cross section ratio for
deep-inelastic positron scattering off nitrogen and helium-3
with respect to deuterium. These 
ratios are compared to similar ratios measured in deep-inelastic 
scattering by NMC \cite{NMC}, E665 \cite{E6652}, and SLAC \cite{SLAC2}. 
The ratio of the inclusive DIS cross sections on
$^{14}$N ($^3$He) and $^2$H is presented in figure~\ref{fig:14n-x}.
A significant difference between the present data and
previous data 
is observed for $x <$ 0.06. In this domain the
HERMES data for both nuclei are smaller than the
NMC and E665 data and the deviation increases towards
smaller values of $x$. At high
values of $x$ the HERMES data 
are in agreement with the SLAC data. In the following
it is shown how the difference between the NMC and HERMES
measurements can be attributed to an $A$-dependence of $R$ at low 
values of $x$ and $Q^2$.

Apart from the data shown in figure \ref{fig:14n-x}, other data
exist which show a strong reduction of 
$\sigma_A / \sigma_D$ for 0.01 $< x <$ 0.1 
and 0.05 $< Q^2 <$ 1.5 GeV$^2$ that is similar to that
of the HERMES data on $^{14}$N \cite{Franz,FNAL}. 
However, these data were never 
used to study a possible $A$-dependence of $R$, either
because of insufficient statistics \cite{FNAL}, or
because of their limited kinematic coverage \cite{Franz}. 

In deep-inelastic
charged lepton scattering from an unpolarised target,
the double-differential cross section per nucleon can be written 
in the one-photon exchange approximation as
\begin{eqnarray}
    \frac{d^2\sigma}{dxdQ^2} & = &
    \frac{4\pi\alpha^2}{Q^4} \frac{ F_2(x,Q^2)} {x} \times \nonumber\\
    & & \left[
            1-y- \frac{xyM}{2E}  + 
                  \frac{y^2}{2} \left(
            \frac{   1+4M^2 x^2 / Q^2 } {1+R(x,Q^2)} \right)
         \right] \nonumber\\
    & = & \frac{\sigma_{\rm{Mott}}}{E^{\prime} E} 
	  \frac{\pi F_2(x,Q^2)}{x \epsilon}
	  \left[ \frac{1 + \epsilon R(x,Q^2)}{1 + R(x,Q^2)} \right],
\end{eqnarray}
where $y = \nu / E$, $\sigma_{\rm{Mott}}$ represents the cross section for
lepton scattering from a point charge,
and $E$ and $E^{\prime}$ are the initial and final lepton energy, respectively. 
The virtual photon polarisation 
parameter is given by
\begin{equation}
	\epsilon = \frac{ 4(1-y) - \frac{Q^2}{E^2} }
			{4(1-y) + 2y^2 + \frac{Q^2}{E^2} }.
\end{equation} 
The ratio of DIS cross sections from nucleus $A$ and deuterium $D$ (=$^2$H)
is then given by:
\begin{equation}
\frac{\sigma_{A}}{\sigma_{D}}=\frac{F_2^A}{F_2^D}
        \frac{(1+ \epsilon R_A)(1+R_D)}{(1+R_A)(1+\epsilon R_D)},
\end{equation}
where $R_A$ and $R_D$ represent the ratio $\sigma_L / \sigma_T$ for
nucleus $A$ and deuterium.
For $\epsilon \rightarrow$ 1 the cross section ratio equals the
ratio of structure functions $F_2^A / F_2^D$.
For smaller values
of $\epsilon$ the cross section ratio is equal 
to $F_2^A / F_2^D$ only if $R_A = R_D$. 
A difference between $R_A$ and $R_D$ will thus
introduce an $\epsilon$-dependence of $\sigma_{A}/\sigma_{D}$.
Hence, measurements of $\sigma_{A}/\sigma_{D}$ as a function of
$\epsilon$ can be used to extract experimental information 
on $R_A / R_D$, if $R_D$ is known.

The data presented in this paper
were collected by the HERMES experiment
at DESY using $^1$H, $^2$H, $^3$He, and
$^{14}$N internal gas targets in the 27.5 GeV 
positron storage ring of HERA. 
The target gases were injected
into a tubular open-ended
storage cell inside the positron ring.
The cell provides a 40 cm long target 
with areal densities
of up to 6 $\times$ 10$^{15}$ nucleons/cm$^2$ for $^{14}$N.
The luminosity was measured by detecting Bhabha-scattered
target electrons in coincidence with the scattered positrons, in a pair of 
NaBi(WO$_4$)$_2$ electromagnetic calorimeters. 
Dead times of less than 5\% were observed even at the highest 
luminosities of about 10$^{33}$ nucleons/(cm$^2$s).
Systematic uncertainties in the measurements of the cross section ratios 
were minimized by cycling among different target gases 
every 2 -- 4 hours during part of the data taking.

In the HERMES spectrometer \cite{specpaper}
both the scattered positrons 
and the produced hadrons can be detected and identified within an
angular acceptance $\pm$ 170 mrad horizontally, and 40 -- 140 mrad
vertically.
The trigger was formed from a coincidence between a pair of scintillator

\noindent
hodoscope planes and a lead-glass calorimeter.
The trigger required an energy of more than 3.5 GeV deposited
in the calorimeter, resulting in a typical trigger rate of
100~Hz. Positron identification was accomplished using 
the calorimeter, the second hodoscope, which functioned as a 
preshower counter, a transition radiation detector,
and a threshold gas \u Cerenkov counter. This system provided
positron identification with an average efficiency of 99 \% and a
hadron contamination of less than 1 \%.

Deep-inelastic scattering events were
extracted from the data by imposing constraints on
$Q^2$, $W$ (the invariant mass of the photon-nucleon system),
and $y$. For each event it was
required that $Q^2 >$ 0.3 GeV$^2$, $W >$ 2 GeV and $y <$ 0.85.

As the ratio $\sigma_A / \sigma_D$ involves nuclei with different
numbers of protons, radiative corrections do not cancel in the
ratio. In particular, the radiative processes associated with
elastic and quasi-elastic scattering are different for the two
target nuclei. These radiative corrections have been computed 
using the methods outlined in Ref. \cite{ABA}. 
In the cross section ratio the correction associated with 
elastic (i.e. coherent) scattering from the target nucleus is dominant.

Several input parameters are needed for the calculation of the radiative 
corrections. For the evaluation of the coherent radiative tails, 
the nuclear elastic form factors must be known. Parameterisations of 
the form factors of $^2$H, $^3$He, and $^{14}$N were taken from the 
literature \cite{deut,heli,nitr}. 
For the quasi-elastic tails, the nucleon form factor parameterisation of 
Gari and Kr\"umpelmann \cite{gari} was used. The reduction of the 
bound nucleon cross section with respect to the free nucleon one 
(quasi-elastic suppression) was evaluated using the results of a calculation 
by Bernabeu \cite{berna} for deuterium and the 
non-relativistic Fermi gas model for $^3$He and $^{14}$N \cite{moniz}.
The evaluation of the inelastic higher order QED
processes requires the knowledge 
of both $F_2$ and $R$ over a wide range of $x$ and $Q^2$. 
The structure function $F_2^D(x,Q^2)$ was described by a parameterisation 
of the NMC, SLAC, and BCDMS data \cite{NMCpar}; for $R_D$  the Whitlow 
parameterisation \cite{R1990} was used. 
As the values of $F_2^A(x,Q^2)$ and $R_A(x,Q^2)$ are unknown for
$^3$He and $^{14}$N,
an iterative procedure has been used. As a starting point
the nuclear structure 
functions $F_2^A(x,Q^2)$ were taken from phenomenological fits to the 
SLAC and NMC data, and $R_A(x,Q^2)$ was assumed to be equal to $R_D(x,Q^2)$.
The resulting radiatively corrected values of $\sigma_A / \sigma_D$ were
used to determine $F_2^A/F_2^D$ and $R_A/R_D$, which were given as input
to the radiative correction code in the next step until convergence was 
reached. It is noted that the large difference between the NMC/E665 and HERMES
values of $\sigma_A / \sigma_D$ 
is already present if the NMC and SLAC parameterisations are used for
$F_2(x,Q^2)$ and $R(x,Q^2)$.
The iteration procedure, which converges in three steps,
enlarges the difference by about 40 \% (for $^{14}$N)  in the lowest
$x$ bins.

The size of the radiative corrections is largest in the lowest
$x$-bin, where it amounts to 0.552, 0.461, and 0.372 for $^{2}$H,
$^{3}$He, and $^{14}$N, respectively.
The systematic uncertainty in the radiative corrections was estimated
by using upper and lower limits of the parameterisations,
or alternative parameterisations \cite{NMCpar,R1990,diff,ArDay}
for all the above input parameters.
The total systematic uncertainty in the cross section ratios 
varies
from 5 \% (4 \%) at low $x$ to 2 \% (1 \%) at high $x$ for $^{14}$N 
($^3$He). It includes the normalization uncertainty of 2 \% (1 \%) and 
the uncertainty in the radiative corrections, which is dominant.

The effects originating from the finite resolution of 
the spectrometer and from the hadron contamination in the positron sample 
have been determined and found to be negligible.
As a cross check of the understanding of the entire
analysis chain including the 
radiative corrections, the cross section ratio of deuterium and hydrogen 
has been determined as a function of $x$ and $Q^2$. The

\begin{figure} [t]
\begin{center}
\includegraphics[width=0.47\textwidth]{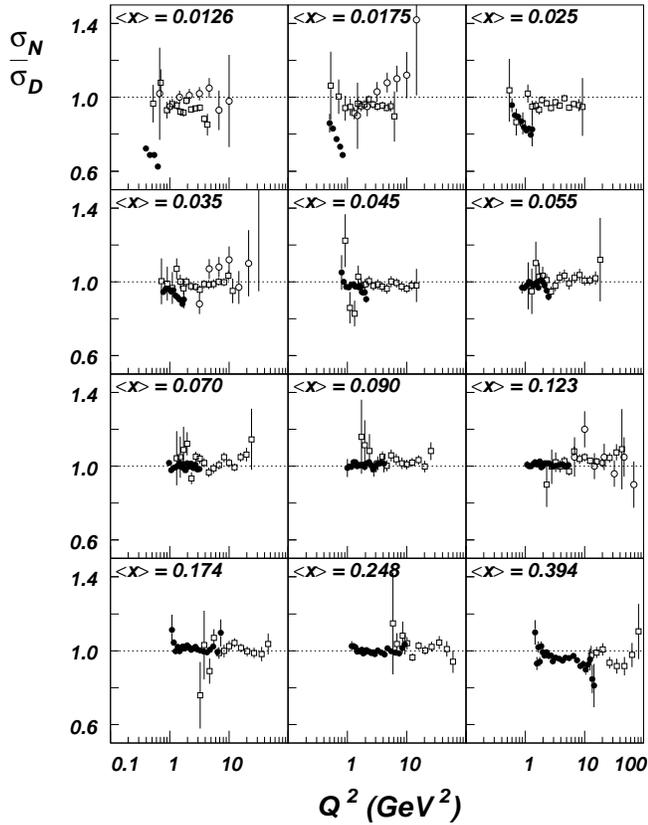}
\epsfxsize=3in
\epsfxsize=\hsize
\end{center}
\caption[]{\label{fig:14n-q2}
Cross section ratios of inclusive deep-inelastic lepton scattering 
from $^{14}$N to $^2$H versus $Q^2$ for specific $x$-bins (solid
circles). Also shown are the $^{12}$C/$^2$H data of NMC (open squares)
and E665 (open circles). Only statistical errors are shown.}
\end{figure}

\noindent
HERMES measurement \cite{Antje} of $\sigma_D/\sigma_p$ agrees within the
systematic uncertainties (3 \% at low $x$ down to 1.5 \% at high $x$) with the
results from earlier experiments \cite{nmcf2np,slacf2np}.

The results of the analysis \cite{jjvh,shin}
are shown in figure \ref{fig:14n-x} as a
function of $x$. It is noted that throughout the analysis
the $\sigma_{3He}/\sigma_D$ data
have been corrected for the excess of protons over neutrons in $^3$He
using the measured $\sigma_D/\sigma_p$ ratios \cite{nmcf2np}.
The $\sigma_{14N}/\sigma_D$ data are displayed in more detail in
figure \ref{fig:14n-q2}
as function of $Q^2$ for fixed values
of $x$. In the first four $x$-bins
a striking discrepancy between the HERMES and NMC data
is observed. The discrepancy
increases with $Q^2$, but at the same time the average deviation
in each $x$ bin decreases with $x$. Moreover, as the data
show a discontinuity with respect to $Q^2$ for the same four
lowest $x$-bins, it is unlikely that the discrepancy observed
in figure \ref{fig:14n-x} is caused by scaling violations of
the ratio $F_2^A(x) / F_2^D(x)$.

\begin{figure} [t]
\begin{center}
\includegraphics[width=0.47\textwidth]{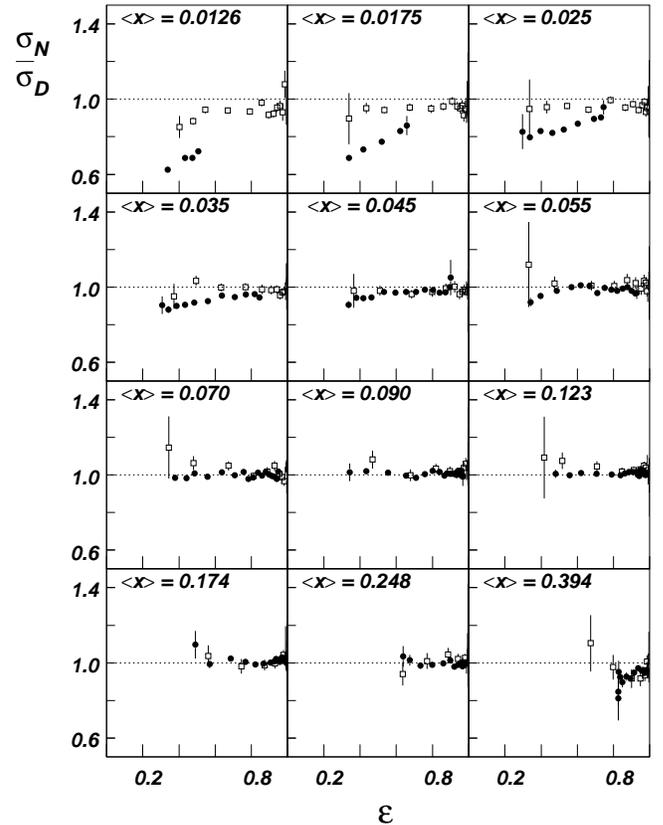}
\epsfxsize=3in
\epsfxsize=\hsize
\end{center}
\caption[]{\label{fig:14n-e}
Cross section ratios of inclusive deep-inelastic lepton scattering 
from $^{14}$N to $^2$H versus $\epsilon$ for individual $x$-bins 
(solid circles). Also shown are the $^{12}$C / $^2$H data of NMC 
(open squares). Note that the two data sets correspond to 
different $Q^2$ ranges. Only statistical errors are shown.}
\end{figure}

As the structure function ratio $F_2^A/F_2^D$ depends only 
on $x$ and $Q^2$, the observed 
difference in the cross section ratios measured at HERMES and NMC/E665 
can be explained only by an $A$-dependence of the ratio $R(x,Q^2)$. 
Therefore, in figure \ref{fig:14n-e}
the data have been plotted
versus $\epsilon$ for fixed values of $x$.
The $\epsilon$-dependence of the HERMES 

\noindent
data shows that the
deviation with respect to unity decreases with
increasing $\epsilon$ values. This is in accordance
with the anticipated behaviour of $\sigma_A / \sigma_D$
if $R_A$ differs from $R_D$, as displayed by Eq. (3).
In contrast to the HERMES data, the NMC data show no or very little
$\epsilon$-dependence. However, the two data sets shown in
figure \ref{fig:14n-e} at the same $\epsilon$ and average $x$
values correspond to different $Q^2$ values, as can be seen by
comparing the same $x$-bins in figures \ref{fig:14n-q2} 
and \ref{fig:14n-e}.

The data of the individual $x$-bins of figure \ref{fig:14n-e} have
been fitted using Eq. (3). Separate fits for the HERMES and
NMC data have been performed. In these fits a parameterisation of
$R_D$ \cite{SLAC} has been used, while the ratios
$R_A / R_D$ and $F_2^A / F_2^D$ have been treated as
free parameters.
A single value of $R_A / R_D$ and $F_2^A / F_2^D$ has been extracted
from each $x$-bin in figure \ref{fig:14n-e} for both the HERMES and NMC
data at the average $x$ and $Q^2$ values of each experiment. In this
procedure it is assumed that
both $R_A / R_D$ and $F_2^A / F_2^D$ are
constant over the limited $Q^2$ range covered by the data in each $x$-bin.
While the $Q^2$-dependence of $F_2^A/F_2^D$ is known to be very
small \cite{NMC-C,SnC}, the effect of a possible 
$Q^2$-dependence on the extracted values of $R_A / R_D$ has been
studied separately. Assuming 

\begin{figure} [t]
\begin{center}
\includegraphics[width=0.47\textwidth]{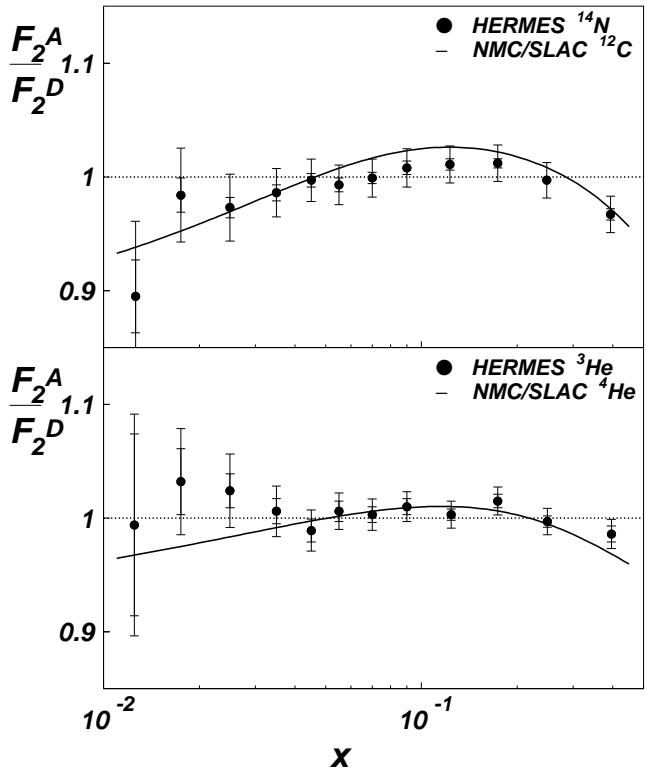}
\epsfxsize=3in
\epsfxsize=\hsize
\end{center}
\caption[]{\label{fig:14n-F2}
The ratio of the inclusive structure functions $F_2^A / F_2^D$ versus $x$
as derived from the fit of the $\epsilon$ dependence of the cross section
ratios. The HERMES data on $^{14}$N are shown in the upper panel and those
collected on $^3$He in the lower panel. The data are compared to 
parameterisations of the $F_2^A / F_2^D$ ratio
for A=4 and 12, which were determined using the NMC 
and SLAC data \cite{NMC,SLAC2}. The inner error bars include the
statistical uncertainty and the correlated error in $R_A / R_D$. The
outer error bars represent the total uncertainty including the
systematic uncertainty.
}
\end{figure}

\noindent
a linear $Q^2$-dependence of $R_A/R_D$,
it has been verified that the
$Q^2$-range covered by each $x$-bin does not affect the average
values of $R_A/R_D$ derived from the fit.

The values of $F_2^A/F_2^D$ derived from the fit of the HERMES data
are displayed in figure \ref{fig:14n-F2} for both nitrogen and
helium-3. The statistical uncertainty and the effect of the
correlated error in $R_A/R_D$ (inner error bars) are indicated
separately from the total uncertainty that includes
the systematic uncertainties as well (outer error bars). 
The data are compared to curves representing
parameterisations of the $F_2^A/F_2^D$ ratios for $A$ = 4 and 12, which
were determined using the NMC and SLAC data \cite{NMC,SLAC2}. The data
are seen to
be in agreement with the parameterisations. The uncertainties on
$F_2^A/F_2^D$ are too large
to observe any systematic deviation between $F_2^A/F_2^D$
for neighbouring nuclei.

\begin{figure} [t]
\begin{center}
\includegraphics[width=0.47\textwidth]{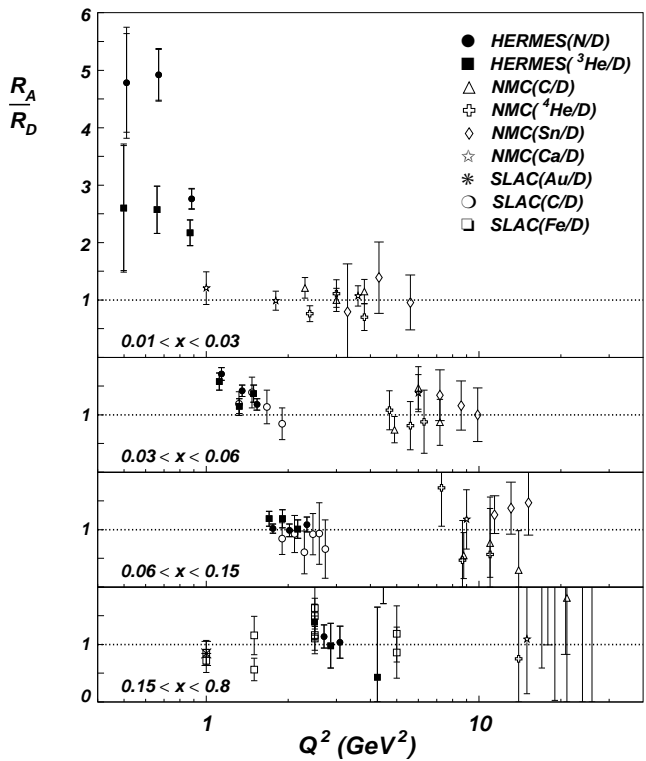}
\vspace{3mm}
\caption[]{\label{fig:rq-all}
The ratio $R_A / R_D$ for
nucleus A and deuterium as a function of $Q^2$ for four different
$x$ bins. The HERMES
data on $^{14}$N ($^{3}$He) are represented by the solid circles (squares).
The open triangles ($^{12}$C) and crosses ($^{4}$He)
have been derived from the NMC data using the same technique. The other
SLAC \cite{AuD} and NMC data \cite{SnC} displayed have been derived from
measurements of $\Delta R = R_A - R_D$ taking a parameterisation
\cite{SLAC} for $R_D$. The inner error bars include both the statistical
uncertainty and the correlated error in $F_2^A/F_2^D$. The outer
error bars also include the systematic uncertainties.
}
\end{center}
\end{figure}

The resulting values of $R_A / R_D$ are shown in figure \ref{fig:rq-all}.
The error bars include the statistical uncertainty, 
the correlated error in $F_2^A/F_2^D$ and the systematic uncertainty.
Both the present results, which were obtained from the fits of the
HERMES and NMC data shown in figure \ref{fig:14n-e}, and the results
derived from other sources are displayed.
The data are plotted at the average value of $Q^2$ for a given $x$-bin.
The values of $R_A / R_D$ derived from the HERMES
data are considerably larger
than unity for $x <$ 0.06 and $Q^2 <$ 1.5 GeV$^2$.
The deviation from unity is smaller for $^3$He than for $^{14}$N, as
one would expect for a medium dependent effect.
The results of the fits of the NMC data for $^4$He and $^{12}$C,
which were collected at higher average $Q^2$ values,
are all consistent with unity.
The present data for $R_A / R_D$ are also compared to
the results of previous studies of the $A$-dependence of $R$.
Existing data are usually represented in terms
of $\Delta R = R_A - R_D$.
The published values of $\Delta R$ \cite{SLAC,AuD,SnC}
have been converted to $R_A /R_D$
using a parameterisation for $R_D$ \cite{SLAC},
and added to figure \ref{fig:rq-all}.
All data above $Q^2$ = 1.5 GeV$^2$ are seen to be 
consistent with unity, while for $x <$ 0.06 and
$Q^2 <$ 1.5 GeV$^2$ a strong $Q^2$-dependence is observed.
It is noted that the $R_A /R_D$ data below $x$ = 0.15
can be described by an inverse power of $Q^2$,
independent of $x$.

Possible mechanisms that give rise to an enhancement
of $R_A$ with respect to $R_D$ at low $Q^2$ and $x$ are
constrained by the present data to not significantly change
the ratio of structure functions $F_2^A / F_2^D$.
Since $F_2^A$ and $R_A$ depend differently on
the longitudinal and transverse DIS 
cross sections, the different effects of the nuclear medium
on $\sigma_L$ and $\sigma_T$ can be distinguished.
In fact, both an enhancement of the longitudinal response and 
a corresponding reduction
of the transverse response are needed to explain the present
data. Evaluated explicitly for the lowest three $x$-bins of
the $^{14}$N/$^2$H data, we find values of 2.15(40), 2.32(25), 
and 1.78(15) for $\sigma_L^A/\sigma_L^D$, and the values
0.45(4), 0.47(3), and 0.65(2) for $\sigma_T^A/\sigma_T^D$.
These results appearing at low $Q^2$ and low $x$ seem to indicate
a novel large shadowing effect not contained in current theoretical
models. 
Possible shadowing differences between $\sigma_L$
and $\sigma_T$ have been discussed in refs. \cite{Niko91,Baro96}.
The quoted enhancement of $R_A$ with respect to $R_D$
does not exceed 50 \% \cite{Baro96}. However, these studies
did not cover the kinematics of the present experiment.
It is noted that a satisfactory description of our data
should also encompass the real-photon data, which show less
shadowing ($\sigma_T^A/\sigma_T^D$ = 0.77(5) for $^{12}$C at
$Q^2$ = 0) than the virtual-photon data at low $Q^2$
(see ~\cite{DESY77,Franz}).

The steep $Q^2$-dependence of the data also suggests an
explanation in terms of a higher twist effect \cite{Ellis82}, 
i.e. strong quark-gluon correlations that are enhanced in the 
nuclear environment \cite{Bars90}. This conclusion is
supported by the fact that leading twist effects are estimated
to be much smaller than the observed enhancement 
of $R_A/R_D$ \cite{BS,Gou96}. The size of twist-4 effects in
nuclei has been estimated by Luo, Qiu and Sterman \cite{LQS}
for dijet photoproduction. They find sizable enhancement
factors of order 100 \% that scale as $A^{1/3}$.  

In order to arrive at
a proper interpretation of the data presented in this paper,
the quoted effects must be 

\noindent
evaluated explicitly for the
conditions of our experiment. Experimentally, it is important
to extend the present measurements to heavier nuclei such that
the $A$-dependence of the effect can be determined precisely.

In summary, deep-inelastic positron scattering data on $^2$H,
$^3$He, and $^{14}$N are presented. At low values of $x$ and
$Q^2$, a large difference is observed between
the presently measured cross section ratios $\sigma_{3He}/\sigma_D$
and $\sigma_{14N}/\sigma_D$, and those reported by previous experiments
at higher values of $Q^2$. Values for the ratio of
$R_A / R_D$ with $R$ the ratio $\sigma_L / \sigma_T$ of longitudinal
to transverse DIS cross sections have been derived
from the dependence of the data on the virtual photon
polarization parameter $\epsilon$. The
data show a strong $Q^2$-dependence of $R_A / R_D$
at low $x$ and $Q^2$ and represent the first observation of a
nuclear effect in the ratio of longitudinal to transverse
photoabsorption cross sections. 
The uncertainty in the measurements is dominated by our estimate
of the systematic uncertainty in the radiative corrections.
In the absence of explicit calculations, it 
can be speculated that our result represents evidence
for the existence of enhanced quark-gluon correlations in atomic nuclei.

\begin{acknowledgements}
We gratefully acknowledge the DESY management for its support and
the DESY staff and the staffs of the collaborating institutions.
This work was supported by
the FWO-Flanders, Belgium;
the Natural Sciences and Engineering Research Council of Canada;
the INTAS and TMR network contributions from the European Community;
the German Bundesministerium f\"ur Bildung, Wissenschaft, Forschung
und Technologie; the Deutscher Akademischer Austauschdienst (DAAD);
the Italian Istituto Nazionale di Fisica Nucleare (INFN);
Monbusho International Scientific Research Program, JSPS, and Toray
Science Foundation of Japan;
the Dutch Foundation for Fundamenteel Onderzoek der Materie (FOM);
the U.K. Particle Physics and Astronomy Research Council; and
the U.S. Department of Energy and National Science Foundation.
\end{acknowledgements}

%%%%%%%%%%%%%%%%%%%%%%%%%%%%%%%%%%%%%%%%%%%%%%%%%%%%%%%%%%%%%%%%%%%%%%%%%

\end{document}